\documentclass[%
 reprint,
 superscriptaddress,
 showpacs,
  nofootinbib,
 amsmath,amssymb,
 aps,
   prl,
]{revtex4-1}

\usepackage{verbatim}
\usepackage{graphicx}
\usepackage{epsfig}
\usepackage{amsfonts}
\usepackage{amsmath,amssymb}
\usepackage{bm}
\usepackage{hyperref}
\usepackage{color}

\begin{document}

\title{Nucleon-nucleon scattering with the Complex Scaling Method and realistic interactions}

\author{G. Papadimitriou}
\email{georgios@iastate.edu}
\affiliation{
Department of Physics and Astronomy, Iowa State University, Ames, Iowa 50011-3160, USA
}%

\author{J. P. Vary}
\email{jvary@iastate.edu}
\affiliation{
Department of Physics and Astronomy, Iowa State University, Ames, Iowa 50011-3160, USA
}%


\begin{abstract}

We demonstrate the validity of the complex scaling method for realistic strong, non-local, nucleon-nucleon interactions by comparing the deuteron bound state and nucleon-nucleon scattering phase shifts with results from other high-precision methods.  
This opens a pathway for the simultaneous \textit{ab initio} solutions of the nuclear bound and scattering problems within a unified framework.

\end{abstract}


\maketitle



{\it Introduction.} One of the most important theoretical endeavors of modern Nuclear Physics (NP) is to develop methods that
address on equal footing structure and scattering observables using the same realistic nuclear Hamiltonian.
This is the only way to consistently understand 
nuclear phenomena sensitive to continuum effects
and also to avoid computing model-dependent quantities such as spectroscopic factors and effective single particle energies \cite{duguet}.
One important region of the nuclear chart that structure and scattering aspects 
overlap is located close to the drip lines, where the nucleons are very weakly bound
or slightly unbound forming a resonant nuclear system.  These exotic systems have implications for nuclear astrophysics.  They are characterized by a very low level density of states,
with few or no bound states, where statistical assumptions that 
underlie some traditional theoretical approaches are less reliable. We show the complex scaling method (CSM), used successfully in Quantum Chemistry, is a viable approach to 
strong nuclear interaction systems.  
Therefore, the CSM will complement existing methods that include the nuclear 
continuum  \cite{Nollett,Navratil1,gaute_michel,volya,review_GSM,Luu_2010,ncgsm,Myo20141} with the advantage that the CSM may be combined with a variety of 
bound state techniques.  
This combination is particularly important for exploiting recent advances in high-performance computing.

Solving the many-body nuclear scattering problem is a very formidable task.
For an exact treatment one solves the scattering equations in either momentum or coordinate 
space. Then, in order to describe resonance features and also access several elastic and inelastic channels the equations are solved
over a  range of energies. 
Coordinate space methods include the Faddeev-Yakubovsky, Hyperspherical  Harmonics using the Kohn variational principle, as well as the Alt-Grassberger-Sandhas equations in momentum space. Especially the latter, after the Coulomb singularity \cite{screening} issue was resolved, is the most
tractable route among the exact methods, since the  scattering boundary conditions are 
naturally imposed when working in momentum space. 
All of the above methods provide precision results for the description of three-nucleon scattering
using realistic nucleon-nucleon (NN) plus
three-nucleon (3N) interactions\cite{Glockle1996107,Viviani,Lazau_FY,reactions_benchm}. Nevertheless, the calculations 
are very involved computationally  \cite{Nogga} making it hard to obtain solutions for systems with
A $\geq$ 4.  Only recently has the four-body scattering problem above the breakup threshold been solved exactly using realistic interactions \cite{Deltuva}.  
The combined mathematical and computational challenge for the imposition of the appropriate scattering boundary conditions in coordinate space techniques, 
the factorial scaling of the  antisymmetrization between the colliding particles, the difficulty of including 
3N interactions 
in momentum space formulations for scattering and also 
the rapid increase of equations one needs to solve in momentum space, 
are the basic bottlenecks for applications of these approaches to heavier systems.
On the other hand bound state techniques such as No-Core Shell Model \cite{Barrett2013131},  
Green's Functions Monte Carlo\cite{GFMC},  Coupled Cluster theory \cite{cc_review}
, In-Medium Similarity Renormalization Group \cite{imsrg},  self-consistent (Gorkov) Green's Functions \cite{scgf} are not limited  by the demands of antisymmetrization. The many-body correlations are well-treated  and  with the increase in computer power they have reached  highly accurate numerical standards. 
Hence it is very important to develop a unification of the bound state and scattering domains by taking full advantage of the 
recent advances in the technologies for solving the bound state domain.

Some of the methods in NP that employ bound state techniques to solve the scattering problem include the Wigner's R-matrix \cite{Descouvemont}, the 
Lorentz Inverse Transformation \cite{Efros1994130},  momentum lattice technique \cite{Kukulin},  continuum discretized coupled-channels \cite{Thompson}  and  CSM \cite{Myo20141,kruppa_george}. 
For a recent review of bound-state techniques for the scattering problem we refer the reader to \cite{Carbonell201455}. For example the CSM was employed to solve
the four body scattering problem above breakup threshold \cite{lazauskas2} with phenomenological 
 nuclear interactions acting only in the S-wave. 
 
In this Communication we describe the bound and scattering problem with the CSM using 
for first time non-local realistic interactions (JISP16) \cite{jisp16} and 
chiral microscopic interactions (N$^3$LO, N$^2$LO$_{opt}$) \cite{EM_pot,n2lo_opt}.  
We apply our technique to the proton-neutron system. We calculate the ground state energy
of the deuteron and  scattering phase-shifts with a single diagonalization without imposing any scattering boundary conditions.
The method itself is a basis expansion technique which employs L$^2$  integrable functions for the description of both the negative and
positive energy spectrum. We show the method with realistic strong and non-local interactions has controllable precision 
which portends the path to a unified description of structure and scattering in heavier nuclear systems.

{\it Formalism.}  In the CSM the coordinates and momenta of the underlying Hamiltonian are rotated as:
$r \, \to \, re^{i\theta}$ and $p \, \to \, pe^{-i\theta}$. Hence a complex rotated local Hamiltonian, with kinetic energy T and potential V, 
takes the form:
$H(r,\theta) \, = \, e^{-2i\theta}T + V(re^{i\theta})$,
where $\theta$ is a real parameter and the resulting Schr\"{o}dinger equation becomes:
\begin{equation}
\label{shroed_cs}
H(r,\theta)\Psi(r,\theta) = E(\theta)\Psi(r,\theta)
\end{equation}
The Hamiltonian  becomes non-Hermitian and its spectrum contains resonant (bound states and resonances) and non-resonant continuum states. It is then a consequence of the Aguilar-Balslev-Combes (ABC) \cite{abc1,abc2} theorem that resonant states  above the 2$\theta$ line are in general invariant with respect to the
rotation angle $\theta$, while the non-resonant continuum states are distributed along cuts rotated by an angle of 2$\theta$ in the complex energy plane.
The rotation point of the continuum states is associated with a many particle threshold. 
Another consequence of the ABC theorem 
is that
the resonant solutions of the Hamiltonian
have decaying asymptotics, e.g. they behave as bound-states, hence any bound state technique for the solution of \eqref{shroed_cs} could be employed.
For an orthonormal basis ensemble $\phi_i(r)$, such as the Harmonic Oscillator (HO) basis, the many-body solution is approximated as:
$\Psi(r,\theta) = \sum_{i=1}^N C_i(\theta)\phi_i(r)$
and \eqref{shroed_cs} leads to a matrix eigenvalue problem:
$\sum_{j=1}^N  H_{ij}(\theta)C_j(\theta) = EC_j(\theta)$.
The main task is to calculate the complex scaled (CS) matrix elements of
the Hamiltonian in the real HO basis.
The transformation is trivial for the kinetic energy operator and only a phase factor $e^{-2i\theta}$ is introduced.
For the kinetic energy in the HO basis the CS matrix elements are given by the analytical tri-diagonal expressions multiplied by the phase factor.
When the effective potential is local and analytical (e.g. Yukawa, Gaussian), the 
CS matrix elements are also easily calculated, since the complex rotation of the coordinate is equivalent to making 
potential parameters complex \cite{Myo20141} and the matrix elements can be calculated either numerically or even analytically in some cases \cite{kato_kruppa}.
That is also the case for realistic meson-exchange potentials as shown for Argonne $\upsilon$8$^{\prime}$\cite{Horiuchi}. 
The difficulty with such potentials that are characterized by a hard core lies in
their singular nature at short distances, which limits the range of $\theta$ values one may employ. It was demonstrated that results become  unstable
with increasing  $\theta$ \cite{lazau_2005} and the problem needs special numerical techniques \cite{Glockle}.   
\begin{figure}[h!] 
  \includegraphics[width=\columnwidth]{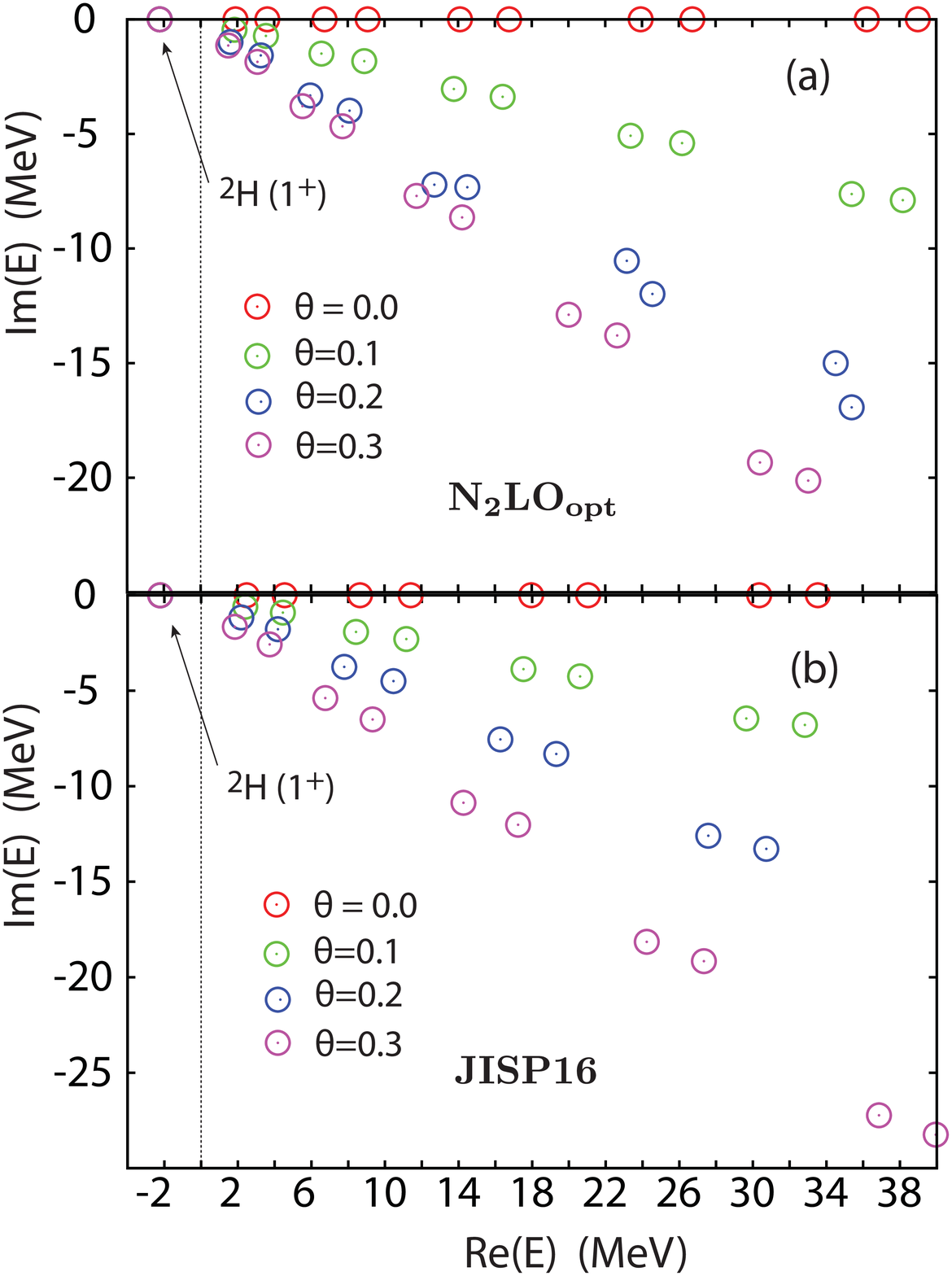} 
  \caption[T]{\label{Fig1}
  (Color online) Eigenvalues of the CS coupled-channel Hamiltonian \eqref{shroed_cs}  for  the N$^2$LO$_{opt}$ (a) and JISP16 (b) realistic potentials.  
  The deuteron bound state (indicated by the arrows) is shown to be invariant under complex rotations of momenta and coordinates.}
\end{figure}
Nowadays, after the successful application of the chiral perturbation theory in NP and also by the application of additional renormalization techniques for the construction
of an effective NN force, the  realistic nuclear potentials are characterized by a softer short range repulsion which may facilitate the use of CSM.
However, these potentials are no longer local and have a complicated structure, which makes the direct application of the CS transformation cumbersome.
In this work we adopt a method first proposed in Refs.\cite{mccur_res,moi_corc} for the calculation of CS matrix elements.
The method involves shifting the CS transformation from the potential to the basis states which we illustrate with a local potential:
\begin{eqnarray}
\label{inv_cs}
\int_{0}^{\infty} \phi_{n}(r)V(re^{i\theta})\phi_{n^{\prime}}(r)r^{2}dr &=& \\ \nonumber
e^{-i3\theta} \, \int_{0}^{\infty} \phi_{n}(re^{-i\theta})V(r)\phi_{n^{\prime}}(re^{-i\theta})r^{2}dr
\end{eqnarray}
where $\phi_n(r)$ are the HO radial basis states which are known analytically. It is straightforward to notice that the CS transformation
on the HO radial basis, corresponds to making the HO length parameter $b$ a complex number, scaled by $b \,  \to \, be^{i\theta}$.
In order to proceed we express our non-local potential operator in terms of relative HO projection operators as:
$V_{b} = \sum_{C,nln^{\prime}l^{\prime}} A_{nln^{\prime} l^{\prime}}^{C;b} |nl,C;b\rangle \langle n^{\prime}l^{\prime},C;b|$ 
where the real numbers $A_{nln^{\prime} l^{\prime}}^{C;b} \, = \, \langle nl,C;b| V | n^{\prime} l^{\prime},C;b \rangle$ are the  relative HO matrix 
elements characterized by the relative quantum numbers
$n$,$l$, the channel $C$ of the interaction (e.g. $^{3}P_0$) and the HO length parameter $b$.
For a general non-local potential in coordinate space we have that:
\begin{eqnarray}
\label{coor_exp}
V(r,r^{\prime}) = \sum_{n,n^{\prime}}A_{nn^{\prime}}\phi_{n}(r)\phi_{n^{\prime}}(r^{\prime})
\end{eqnarray}
where we suppress additional indeces for compactness.
In \eqref{coor_exp} 
we apply the approach of Refs.\cite{mccur_res,moi_corc} to obtain the CS expression:
\begin{equation}
\label{transf_cs}
V^{CS}_{b} = \iint r^2 r^{2 \prime} dr dr^{\prime} \phi_{n}^{be^{i\theta}}(r) V(r,r^{\prime})\phi_{n^{\prime}}^{be^{i\theta}}(r^{\prime})
\end{equation}
The implementation of the CSM reduces
\begin{figure}[h!] 
  \includegraphics[width=\columnwidth]{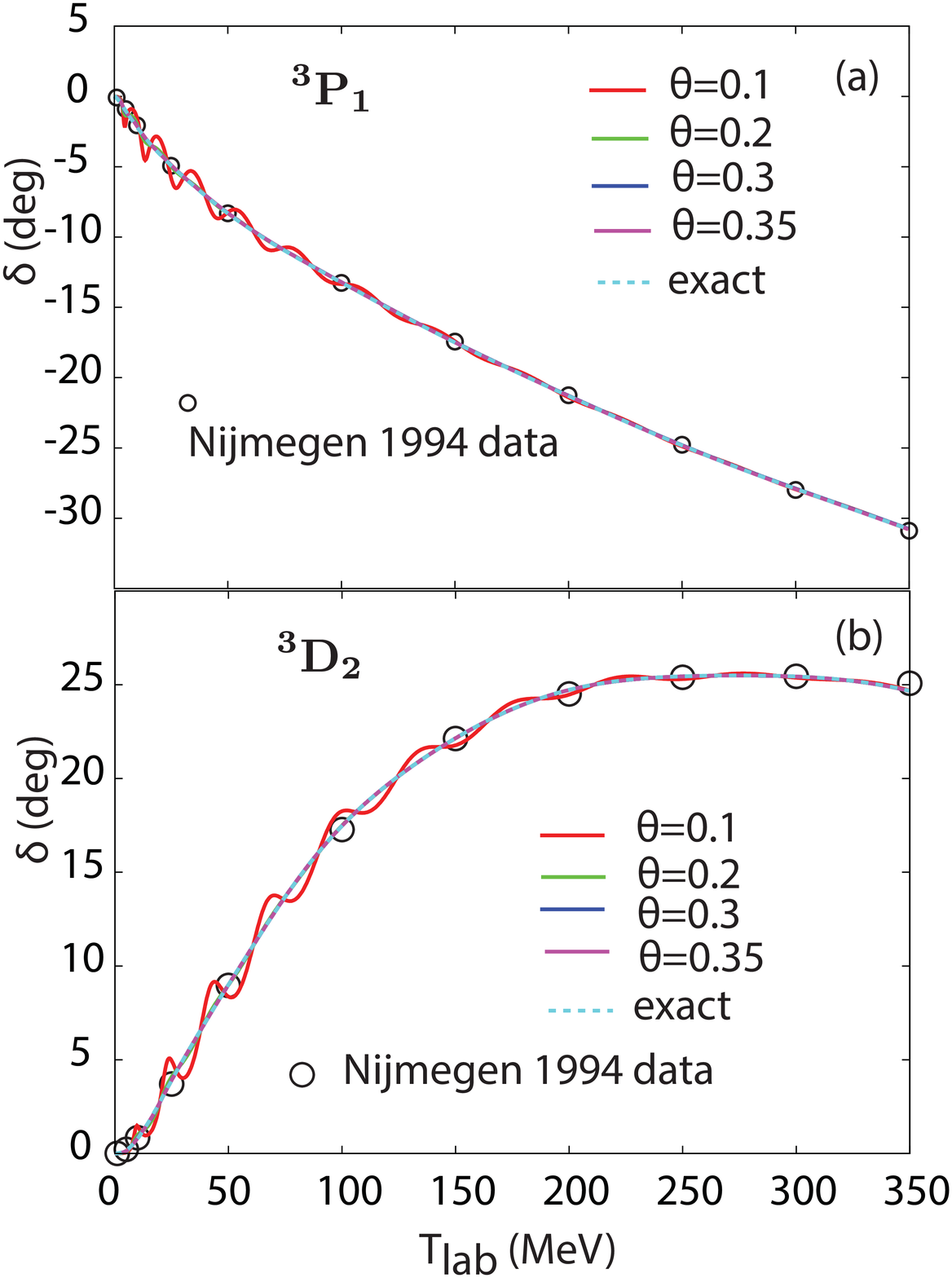}
  \caption[T]{\label{Fig2}
  (Color online) $^3P_1$ (a) and $^3D_2$ (b) channel phase-shifts obtained with the CSM with the JISP16 interaction, as compared with the exact solution and the Nijmegen data
  \cite{Nijmegen}.
  We observe an independence of the results starting already for values of $\theta$ as small as 0.2 rad. Even for $\theta$ = 0.1 rad the phase-shift fluctuations are fairly small and 
  follow the trend of the exact phase-shift.  }
\end{figure}
to a  calculation of this double integral.       
We first tested the validity of this treatment against local Gaussian potentials where we knew both analytical 
and numerical solutions \cite{kato_kruppa}.
In the following, once we have obtained the CS representation of the nuclear potential matrix elements, we diagonalize
\eqref{shroed_cs} for the two-body proton-neutron ($pn$) system.

{\it Results.} 
We first apply the formalism to the ground state (g.s.) of the deuteron ($^{3}$S$_{1}$ - $^{3}$D$_{1}$ coupled channels).
Our goal is to show if the consequences of the ABC theorem, namely the invariance of the bound state with respect to the rotation angle, holds
for the general potentials that we are investigating here. In this demonstration we are using the JISP16 and the N$^2$LO$_{opt}$  chiral interactions at
$b = \hbar /\sqrt{\mu \omega}$ = 1.4399 fm or ($\hbar \omega$ = 40 MeV).
We gather our results in Fig.1. For this calculation we used rotation angles ranging from 0.1 rad to 0.3 rad. We see that the deuteron ground state energy is invariant.
For this application we used the cutoff $N$ = 45 for the number of HO basis states.

Having diagonalized the CS Hamiltonian, in addition to the bound-state(s) and resonanse(s) (if they exist), we also 
obtain the non-resonant scattering continua along the 2$\theta$ rays in the complex plane. It has been shown \cite{Suzuki01062005}  that the 
\begin{figure}[h!] 
  \includegraphics[width=\columnwidth]{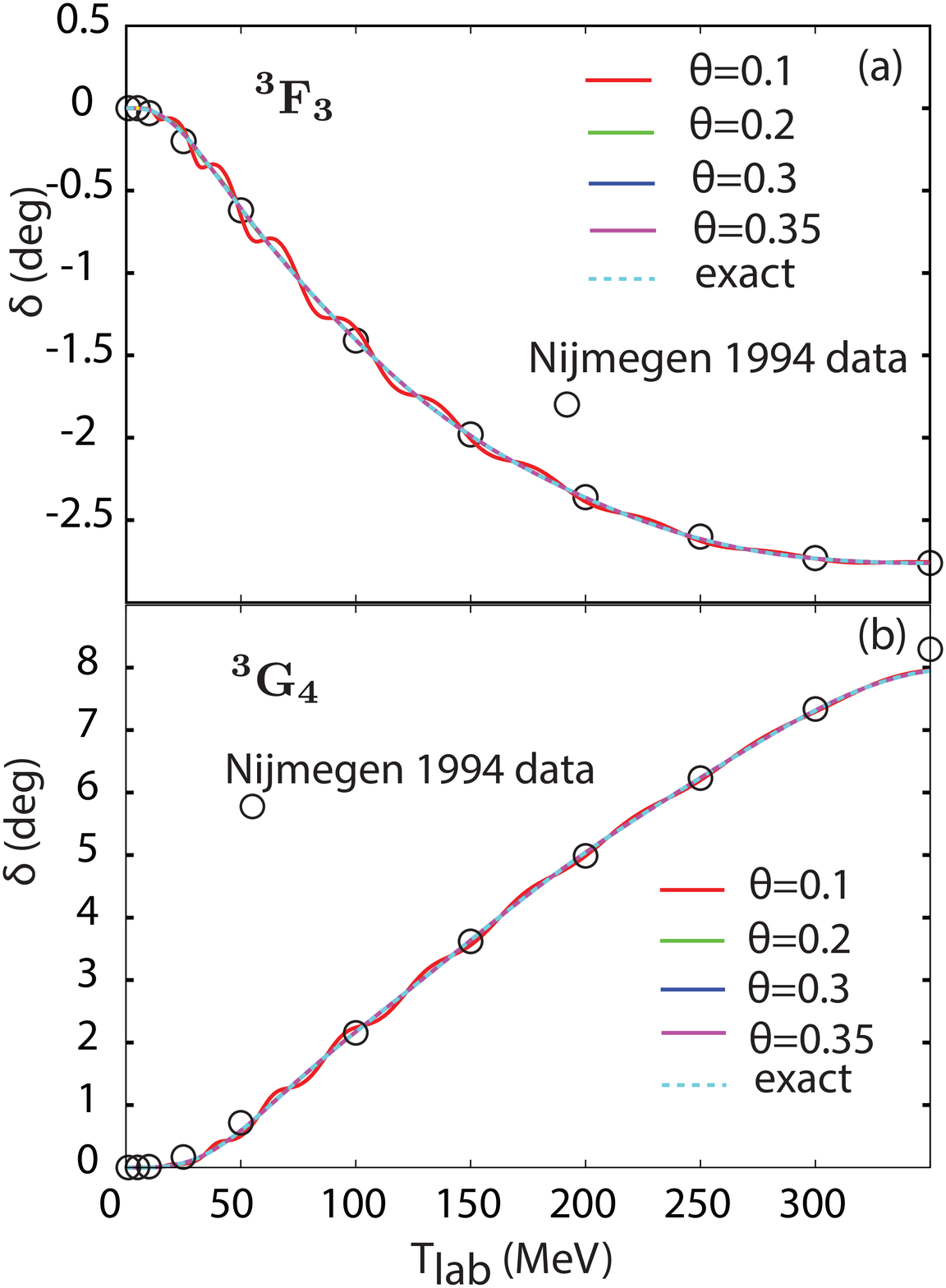}
  \caption[T]{\label{Fig3}
  (Color online) Same as Fig.\ref{Fig2} but for the $^3F_3$ (a) and $^3G_4$ (b) channel phase-shifts.}
\end{figure}
scattering solutions may be used to calculate elastic scattering phase-shifts, by evaluating the CS Continuum Level Density (CLD). The proof is based on the utilization of an
extended completeness relation \cite{Myo01051998,Giraud2003115} which involves resonant and non-resonant scattering states and was originally proposed by Berggren \cite{Berggren}. 
Following \cite{Suzuki01062005}  we calculate
the CLD in the CSM which is defined as:
\begin{equation}
\label{cld_eq}
\Delta^{\theta}(E) = -\frac{1}{\pi}Im \int d{\bf r} \langle {\bf r} |  \frac{1}{E - H^{\theta} } -\frac{1}{E-H^{\theta}_{0}} | {\bf r^{\prime}} \rangle
\end{equation}
where $H^{\theta}_{0}$ is the CS asymptotic part of the Hamiltonian, which in our case is identical to the kinetic energy and $H^{\theta}$
is the total CS interacting Hamiltonian.

The phase shift then is obtained from \eqref{cld_eq} after integrating over the range of energies.
We apply these formulae using our solutions for realistic interactions and we calculate selected uncoupled channel phase-shifts with angular momentum L=1,2,3,4. 
When using the formula \eqref{cld_eq} for the evaluation of the CLD and hence the scattering phase-shift for the
$^1$S$_{0}$ channel, 
we encounter a need, as others have shown\cite{contour_dm}, for more advanced numerical techniques.  The challenge is the very rapid rise in the phase shift at low energy due to 
the virtual (anti-bound) state (pole on the second Riemann sheet) of the $^1$S$_{0}$  channel.
This special case for phase shifts requires larger HO basis sets beyond our current numerical techniques.  However, we would still use the standard procedure to identify a true resonance 
without the phase shifts by its complex pole  (eigenvalue of the CS Hamiltonian) 
that is stable with increasing $\theta$, as we saw above for the deuteron ground state.
As expected, we find no true continuum resonance in the $np$ channels investigated here.

For the orbital angular momentum L $\geq$ 1 uncoupled channels we solve for the phase 
shifts using the CSM plus CLD treatment and compare 
with results of exact calculations using the Schr\"{o}dinger
equation. The results are gathered in Figs.\ref{Fig2},\ref{Fig3}.  For $\theta$=0.1 rad the phase-shifts exhibit small fluctuations which are smeared out
with increasing rotation angle. We notice that results already for $\theta$ = 0.2 rad are 
practically indistinguishable, coincide with 
the exact ones and become independent of the rotation angle $\theta$. 
\begin{figure}[h!] 
  \includegraphics[width=\columnwidth]{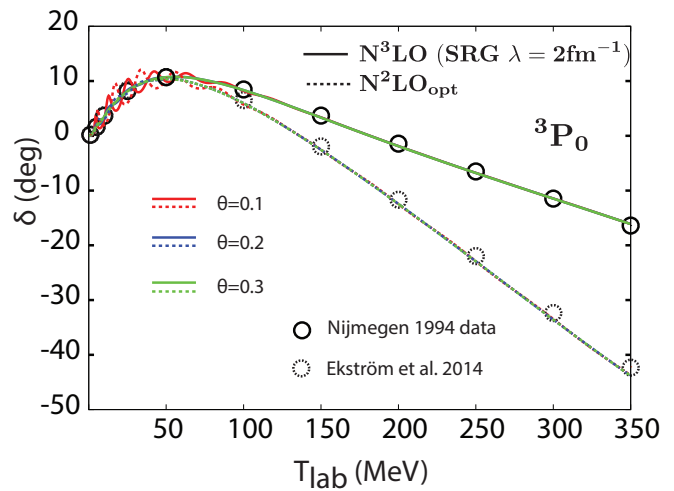}
  \caption[T]{\label{Fig4}
  (Color online) $^3$P$_0$ channel phase shift calculated with the CSM with the N$^3$LO SRG ($\lambda$ = 2.0fm$^{-1}$ and the N$^2$LO$_{opt}$
  realistic interactions.  N$^2$LO$_{opt}$ phase shift data are from \cite{nnlopt})}
\end{figure}
In order to demonstrate the general nature of our approach we present  
results (Fig.\ref{Fig4}) for the $^3$P$_0$ channel scattering phase shift 
using the Entem-Machleidt N$^3$LO interaction which was evolved via a Similarity Renormalization Group (SRG) \cite{bogner}
transformation at a cutoff $\lambda$ = 2.0 fm$^{-1}$ and also with the bare N$^2$LO$_{opt}$. Similar to the results obtained with the JISP16
interaction, our phase shifts obtained from the CS non-resonant scattering solution of \eqref{shroed_cs} show a fast convergence
with respect to the CSM rotation angle variations and already for $\theta$ = 0.2 rad the fluctuations are diminished. 
For this calculation we used $N$ = 35. The JISP16 potential HO basis matrix elements are characterized 
selecting, in Eq.\eqref{transf_cs}, a value $b$ = 1.4399 fm ($\hbar \omega$ = 40 MeV),
while $b$ = 1.5178 fm ($\hbar \omega$ = 36 MeV) for N$^3$LO, and
$b$ =  1.6099 fm ($\hbar \omega$ = 32 MeV) for N$^2$LO$_{opt}$.  
We note that for the phase-shifts calculation using the CLD formulas we limited ourselves to uncoupled channels.
In general, the evaluation of phase-shifts for coupled-channels within the CSM could be feasible by either of two routes.
We may follow the path of Suzuki {\textit et al} \cite{Suzuki_cc_cld} where
the authors defined the CLD in a matrix form for coupled channels. 
One then diagonalizes the CLD matrix and takes the eigenvalues as the partial level densities in a specific eigenchannel. 
Having obtained the partial level density we hope that one could apply formulas similar to the ones used in the uncoupled case for the phase-shifts, even though Suzuki
et al warn that this may be problematic.
An alternative route would be to use the formalism of \cite{kruppa_scat_ampl} and \cite{lazauskas1} which resembles the
initial work of Nuttall and Cohen \cite{Nuttall_Cohen} and was applied for 
the calculation of scattering amplitudes and phase-shifts in a coupled channel case using the CSM. 
Then we may employ the technology we developed for treating realistic non-local interactions within the CSM.

{\it Conclusions.}
The CSM is a state-of-the-art method which truly unifies structure and scattering problems. 
It is a bound state technique which eliminates the need to impose boundary condition for the scattering problem.
We applied the CSM to the $pn$ system and demonstrated numerically the validity of the 
ABC theorem for a general class of potentials (i.e. local potentials were assumed for the proof of the ABC theorem).
Specifically, we showed the invariance with respect to the rotation angle 
$\theta$ for the deuteron g.s. with realistic, even nonlocal, NN interactions for the first time.
In addition, using the solutions we obtained from the CSM,
we calculated single channel scattering phase-shifts by discretizing the continuum
in a HO basis and evaluating the continuum level density. 
The convergence of the phase-shifts as a function of the CS rotation angle parameter
is rather rapid.
The success of the CSM is tied to the fact that the modern nuclear interactions (either chiral or phenomenological) are characterized by 
a manageable short-range repulsion which eliminates possible numerical instabilities appearing in older realistic force applications of CSM 
(with strictly local and analytical forces) in NP. 
Using the HO basis for our calculations one may transform the Hamiltonian matrix elements, 
including 3N interactions, into the laboratory frame and use current many-body solvers.
The fact that we are able to apply this powerful technique without limiting the type of the potential, 
opens a window for more reliable calculations of exotic nuclei and enables assess 
to energies below threshold, resonance parameters  and scattering observables, within a unified approach.

\begin{acknowledgements}
Discussions with A.T. Kruppa, G. Hagen on CSM with realistic interactions matters and 
P. Maris and C. Yang on the diagonalization of 
 complex matrices using numerical libraries such as ARPACK 
are gratefully acknowledged. We would like to thank C. W. McCurdy for sending us useful 
references to CSM in Quantum Chemistry.
This work was supported by the US DOE under grants No. DESC0008485 (SciDAC/NUCLEI) and DE-FG02-87ER40371.
\end{acknowledgements}

\bibliographystyle{apsrev4-1}
\bibliography{CS_real}    

\end{document}